\documentclass[preprint,showpacs,preprintnumbers,amsmath,amssymb]{revtex4}

\usepackage{graphicx}
\usepackage{dcolumn}
\usepackage{bm}
\usepackage{epsfig}

\begin{document}


\title{Thermal and magnetic properties of spin-$1$ magnetic chain  
compounds with large single-ion and in-plane anisotropies}

\author{M.T. Batchelor, Xi-Wen Guan and Norman Oelkers}
\affiliation{Department of Theoretical Physics, Research School of Physical Sciences and Engineering\\
and Mathematical Sciences Institute,\\
Australian National University, Canberra ACT 0200,  Australia}

\date{\today}

\begin{abstract}
\noindent
The thermal and magnetic properties of spin-$1$ magnetic chain  
compounds with large single-ion and in-plane 
anisotropies are investigated via the integrable $su(3)$  model in terms of the quantum transfer 
matrix method and the recently developed high temperature expansion method for exactly solved models. 
It is shown that large single-ion anisotropy may result in a singlet gapped phase in the spin-$1$ 
chain which is significantly different from the standard Haldane phase.
A large in-plane anisotropy may destroy the gapped phase.
On the other hand, in the vicinity of the critical point
a weak  in-plane anisotropy leads to a different phase transition than the Pokrovsky-Talapov transition.
The magnetic susceptibility, specific heat and magnetization 
evaluated from the free energy are in excellent agreement with the experimental data 
for the compounds Ni(C$_2$H$_8$N$_2$)$_2$Ni(CN)$_4$ and Ni(C$_{10}$H$_8$N$_2$)$_2$Ni(CN)$_4$$\cdot$H$_2$O.

\end{abstract}



\maketitle

\section{Introduction}

Haldane's \cite{Hald} conjecture  that spin-$S$ chains exhibit an energy gap in the lowest magnon 
excitation for $2S$ even with no significant gap for $2S$ odd inspired a great deal of 
experimental and theoretical investigation.
Rich and novel quantum magnetic effects, including valence-bond-solid
Haldane phases and dimerized phases \cite{AFF1,LADD1}, fractional
magnetization plateaux \cite{FPL} and spin-Peierls transitions \cite{SPT} have since 
been found in low-dimensional spin systems.
In this light, the spin-$1$ Heisenberg magnets have been extensively studied in Haldane 
gapped materials \cite{SP1C1,SP1C2}. 
The valence-bond-solid ground state and the dimerized state form the Haldane phase 
with an energy gap \cite{AFF1}. 
The Haldane gap in integer spin chains may close in the presence of additional 
biquadratic terms or in-plane anisotropies. 
In particular a large single-ion anisotropy may result in a singlet
ground state \cite{AFF3,Tsvelik} which is significantly different from
the standard Haldane phase. 

The difference between the two gapped
phases appears to arise from the ground state and excitations. In the
Haldane nondegenerate ground state, a single valence bond connects
each neighbouring pair to form a singlet. An expected excitation comes
from breaking down the valence bond solid state where a nonmagnetic
state $S_i=0$ at site $i$ is substituted for a state $S_i=1$. In this
way a total spin $S=1$ excitation causes an energy gap referred to as the
Haldane gap. Whereas the large-anisotropy-induced gapped phase in the
spin-$1$ chain is caused by trivalent orbital splitting. For a large
single-ion anisotropy, the singlet can occupy all states such that the
ground state lies in the nondegenerate gapped phase. The lowest
excitation arises as the lower component of the doublet is involved in the
ground state. This excitation results in the energy gap.

A number of spin-1 magnetic chain compounds have been identified as planar Heisenberg
magnetic chains with large anisotropy.
These include
 Ni(C$_2$H$_8$N$_2$)$_2$Ni(CN)$_4$ (abbreviated NENC),
 Ni(C$_{11}$H$_{10}$N$_2$O)$_2$Ni(CN)$_4$ (abbreviated NDPK) \cite{NENC,sus} and
 Ni(C$_{10}$H$_8$N$_2$)$_2$Ni(CN)$_4$$\cdot$H$_2$O (abbreviated
 NBYC) \cite{NBYC}.
This kind of system exhibits a nondegenerate ground state which can be
separated from the lowest excitation.  
This gapped phase also occurs in some nickel salts with a large zero-field
splitting, such as NiSnCl$_6\cdot 6$H$_2$O \cite{PRB3488},
[Ni(C$_5$H$_5$NO)$_6$](ClO$_4$)$_2$ \cite{PRB3523} and
Ni(NO$_3$)$_2\cdot 6$H$_2$O \cite{PRB4009}.
The theoretical study of these compounds relies on a molecular field approximation 
for the Van Vleck equation \cite{Carlin}. 
To first-order Van Vleck approximation, the exchange interaction is neglected.
To obtain a good fit to the experimental data an effective crystalline field 
has to be incorporated. 
This approximation causes uncertainties and discrepancies in fitting the experimental data. 
Here we take a new approach via the theory of integrable models. 

It recently has been demonstrated \cite{HTE1} that integrable
models can be used to study real ladder compounds via the
thermodynamic Bethe Ansatz (TBA) \cite{TBA} and the exact high temperature
expansion (HTE) method \cite{HTE2,ZT}. 
In this paper we present an integrable 
spin-$1$ chain with additional terms to account for planar single-ion 
anisotropy and in-plane anisotropy.
The ground state properties and the thermodynamics of the chains are
studied via the TBA and HTE.
We show that a large planar single-ion anisotropy results in a 
nondegenerate singlet ground state which is significantly different 
from the Haldane phases found in Haldane gapped materials \cite{SP1C1,SP1C2}.
We examine the thermal and magnetic properties of the compounds 
NENC \cite{NENC,sus} and NBYC \cite{NBYC}.
Excellent agreement between our theoretical results and the
experimental data for the magnetic susceptibility, specific heat and 
magnetization confirms  that the strong single-ion anisotropy, which is
induced by an orbital splitting, can dominate the low temperature
behaviour of this class of compounds.
Our exact results for the integrable spin-$1$ model may provide
widespread application in the study of thermal and magnetic properties
of other real compounds, such as NDPK \cite{NENC,sus} and certain  nickel
salts \cite{PRB3488,PRB3523,PRB4009,Carlin}.

\section{The integrable spin-$1$ model}

In contrast to the standard Heisenberg spin-$1$ materials, experimental 
measurements on the new  spin-$1$
compound LiVGe$_2$O$_6$ \cite{SP1} and the compounds
NENC and NBYC \cite{NENC,NBYC} exhibit unexpected behaviour, 
possibly due to the presence of biquadratic
interaction and a strong single-ion anisotropy, making it very amenable to our approach. 
The axial distortion of the crystalline field 
in the compounds NENC and NBYC results from the triplet $^3A_{2g}$ splitting.
Specifically, the triplet orbit splits into a low-lying doublet ($d_{xy}, d_{yz}$) 
and a singlet orbital ($d_{xz}$) at an energy $\Delta_{CF}$ above the doublet.
Inspired by the high temperature magnetic properties of this kind of material, 
we consider an integrable spin-$1$ chain with Hamiltonian
\begin{eqnarray}
{\cal H}&=&J\,{\cal H}_0+D\sum_{j=1}^N(S_j^z)^2+E\sum_{j=1}^N((S_j^x)^2-(S_j^y)^2) \nonumber\\ 
& &
-\mu_Bg H\sum_{j=1}^N S^z_j, \label{Ham1}\\
{\cal H}_0&=& \sum_{j=1}^{N}\left\{\vec{S}_j\cdot \vec{S}_{j+1}+(\vec{S}_j\cdot
\vec{S}_{j+1})^2\right\}. \nonumber
\end{eqnarray}
${\cal H}_0$ is the standard $su(3)$ integrable spin chain, which is well 
understood \cite{U,BA,Fujii,sun}. 
Here $\vec{S}_i$ denotes the spin-$1$ operator at site $i$, $N$ is the
number of sites and periodic boundary conditions apply. 
The constants $J$, $D$ and $E$ denote exchange spin-spin
coupling, single-ion anisotropy and in-plane anisotropy, respectively. 
The Bohr magneton is denoted by $\mu_B$ and $g$ is the Land$\acute{e}$ factor.
We consider only antiferromagnetic coupling, i.e. $J>0$ and $D>0$.
 
\subsection{The ground state at zero temperature}
\label{sec:TBA}

\begin{figure}
\begin{center}
\includegraphics[width=0.95\linewidth]{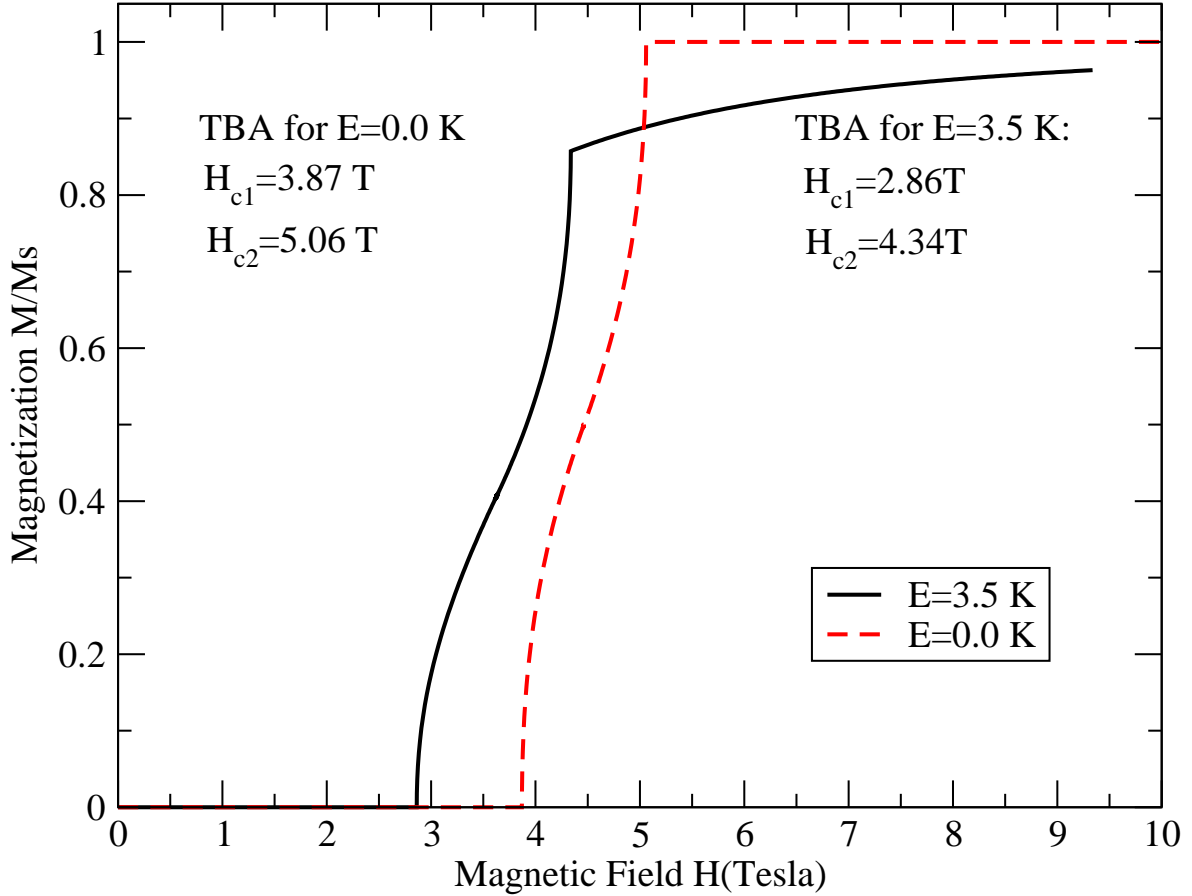}
\caption{Magnetization versus magnetic field $H$ in units of
saturation magnetization for Hamiltonian (\ref{Ham1}) with $J=0.2$ K,
$D=6$ K, $E=3.5$ and $0$ K, $g=2.0$ with parallel
magnetic field. The solid and dashed lines denote the magnetization
derived from the TBA with $E=3.5$ K and $E=0$ K, respectively. The
magnetization curve for $E=3.5$ K indicates different quantum phase
transitions in the vicinity of $H_{c1}$ and $H_{c2}$ than the $E=0$ K square
root field-dependent critical behaviour in the absence of the in-plane anisotropy.
}
\label{fig:SP1SZ}
\end{center}
\end{figure}

For the sake of simplicity in analyzing the ground state properties at zero temperature, 
we first take $E=0$, i.e., no in-plane anisotropy. 
In this case Hamiltonian (\ref{Ham1}), which can be derived from the $su(3)$ row-to-row 
quantum transfer matrix with appropriate chemical potentials in the fundamental basis, 
is integrable by the Bethe Ansatz. 
The energy is given by
\begin{equation}
{\cal E}=-J\sum_{j=1}^{M_1}\frac{1}{(v_j^{(1)})^2+\frac{1}{4}}-DN_0-\mu_BgH(N_+-N_-),
\end{equation}
where the parameters $v_j^{(1)}$ satisfy the Bethe equations \cite{U,BA}
\begin{eqnarray}
\prod_{i=1}^{M_{k-1}}\frac{v_j^{(k)}\!-\!v_i^{(k-1)}\!+\!\frac{\mathrm{i}}{2}}
{v_j^{(k)}\!-\!v_i^{(k-1)}\!-\!\frac{\mathrm{i}}{2}}
&=&
\prod^{M_k}_{\stackrel{\scriptstyle l=1}{l\neq j}}
\frac{v_j^{(k)}-v_l^{(k)}+\mathrm{i}}{v_j^{(k)}-v_l^{(k)}-\mathrm{i}}\nonumber\\
&
 \times & 
\!\! \prod^{M_{k+1}}_{l=1}\!\!\frac{v_j^{(k)\!\!}-\!v_l^{(k+1)}\!-\!\frac{\mathrm{i}}{2}}
{v_j^{(k)}\!\!-\!v_l^{(k+1)}\!+\!\frac{\mathrm{i}}{2}}.\label{BE}
\end{eqnarray}
In the above, $k=1,2$ and $ j=1,...,M_k$ and the conventions
$v_j^{(0)}=v_j^{(3)}=0,\, M_3=0$ apply. $N_{+},N_{0},N_{-}$ denote
the number of sites with spin $S^z=1,0,-1$ in the Bethe eigenstates.
In the thermodynamic limit, the Bethe ansatz equations (\ref{BE})
admit complex string solutions \cite{TBA} from which the TBA equations
can be derived \cite{TBA2, ying}.
Following the standard TBA analysis, we find that the
ground state in the zero temperature limit is gapped if the single-ion
anisotropy $D \!>\! 4J$.  
The singlet ground state is separated from
the lowest spin excitation by an energy gap $\Delta =D-4J$. This
energy gap is decreased by the external magnetic field $H$. 
At the
critical point $H_{c1}=(D-4J)/\mu_Bg$, the singlet ground state breaks
down. Due to the magnon excitation, the magnetization almost linearly
increases with the magnetic field. Once the magnetic field
is increased beyond the second critical point $ H_{c2}=( D+4J)/\mu_Bg$
the ground state is fully polarized, i.e., in the $M\!=\!M_s$ plateau
region. 
The magnetization derived from the TBA is shown in figure \ref{fig:SP1SZ}.
We remark that a gapped phase exists only for
anisotropy values satisfying the `strong anisotropy' condition $D> 4J$. 
As shown in Ref.~\cite{TBA2, ying} for the spin ladders, the magnetization in the
vicinity of the critical fields $H_{c1}$ and $H_{c2}$ depends on the
square root of the field, indicating a Pokrovsky-Talapov transition.
In this regime, the anisotropy effects overwhelm the contribution from the biquadratic
interaction and open a gapped phase in the ground state.

When $E\neq 0$, the in-plane anisotropy $x^2-y^2$ breaks the
$z^2$ symmetry and weakens the energy gap. 
In the presence of the in-plane anisotropy term $E$, the energies split
into three levels with respect to the new basis 
$\phi_0 \!\! = \, \mid \!\!\! 0 \, \rangle$ and 
$\phi_{\pm} \!\! = a_{\pm} \!\!\! \mid\!\!\! -1 \, \rangle + \!\!\! \mid\!\!\!1\, \rangle$, 
with $a_{\pm}=[\mu_BgH\pm \sqrt{(\mu_BgH)^2+E^2}\,]/E$.  
In this basis the eigenvalues of the underlying permutation operator are the same
as the eigenvalues using the fundamental basis.
The model thus remains integrable.
We find that if $E<D$, there is still a gapped phase
with gap $\Delta =D-4J-\sqrt{(\mu_BgH)^2+E^2}$  for the region $H<H_{c1}$. 
Here the critical field
$H_{c1}=\sqrt{(D-4J)^2-E^2}/\mu_Bg$. In this gapped phase the ground
state is the non-degenerate singlet. Subsequently, when $H>H_{c1}$ the
state $\phi_-$ gets involved in the ground state. At the critical
point $H_{c1}$, the phase transition is not  
of the Pokrovsky-Talapov type due to the mixture of a doublet state in
the $\phi_-$ state. 
The magnetization increases as the magnetic field increases.
Past the second critical point
$H_{c2}=\sqrt{(D+4J)^2-E^2}/\mu_Bg$, the singlet state is no
longer involved in the ground state. The state $\phi _-$ fully occupies the
ground state. 
As the magnetic field is increased beyond $H_{c2}$,
the (normalized) magnetization $M={H}/{\sqrt{H^2+(E/\mu_Bg)^2}}$ 
gradually approaches $M_s=1$. 
These novel phase transition may be observed from the low temperature
magnetization curve, which can be evaluated from the TBA equations at
$T=0$, as per the example in figure \ref{fig:SP1SZ}. It shows that the gap
sensitively depends on the single-ion anisotropy and the in-plane anisotropy. 
These phase transitions disappear at high temperatures. In addition,
the inflection point at $H=\sqrt{D^2-E^2}/\mu_Bg$ and
$M=\frac{1}{2}\sqrt{1-({E}/{D})^2}$ indicates that the
probabilities of the components $\phi_0$ and $\phi_-$ are equal.
Moreover, if the exchange interaction
decreases, the magnetization in the vicinity of the critical point
$H_{c1}$ increases steeply. For $J=0$, i.e. the case of independent
spins, the critical points $H_{c1}$ and $H_{c2}$ merge into one
point, at which a discontinuity in the magnetization occurs. For
$D<4J+E$, there is no gapped phase.

\subsection{Magnetic properties at high temperature}

In order to study thermodynamic properties, we adopt the
 Quantum-Transfer-Matrix (QTM) approach \cite{QTM}.  Explicitly,
 following \cite{ZT} the eigenvalue of the QTM for the model
 (\ref{Ham1}) (up to a constant) is given by
\begin{eqnarray}
T^{(1)}_1(v,\left\{v^{(a)}_i\right\})&=&e^{\beta \mu_1}\phi _-(v-\mathrm{i})
\phi _+(v)\frac{Q_1(v+\frac{\mathrm{i}}{2})}{Q_1(v-\frac{\mathrm{i}}{2})}\nonumber\\
&+&
e^{\beta \mu_2}\phi _-(v)\phi _+(v)
\frac{Q_1(v-\frac{3\mathrm{i}}{2})Q_2(v)}{Q_1(v-\frac{\mathrm{i}}{2})Q_2(v-\mathrm{i})}\nonumber\\
&+&e^{\beta \mu_3}\phi _-(v)\phi _+(v+\mathrm{i})
\frac{Q_2(v-2\mathrm{i})}{Q_2(v-\mathrm{i})}.
\label{EQTM1}
\end{eqnarray}
  
In the above equation the chemical potential terms are
\begin{equation}
\mu_1=\mu_BgH,\,\,\mu_2=D,\,\, \mu_3=-\mu_BgH,
\label{cp1}
\end{equation}
where for the moment we take $E=0$.
We have adopted the notation from \cite{ZT} with $\phi _{\pm}(v)=(v\pm
\mathrm{i}u_N)^{\frac{N}{2}}$,
$Q_a(v)=\prod_{i=1}^{M_a}(v-v_i^{(a)})$ for $a=1,2$, and
$Q_0(v)=1$.  Here $u_N=-J\beta/N$ where $N$ is the Trotter
number. Following the HTE scheme \cite{ZT}, we derive the high
temperature expansion for the free energy of model (\ref{Ham1}) in
powers of ${J}/{T}$. 
Because the expansion parameter ${J}/{T}$ is small for weak intrachain coupling $J$, 
we may expect the free energy to accurately describe the thermodynamic quantities at 
sufficiently high temperatures, even for a small number of terms. 
To third order, the result is
\begin{eqnarray}
-\frac{1}{T}f(T,H)&=&\ln{C_0} + C^1_{1,0}\, \frac{J}{T} 
+ C^1_{2,0} \left(\!\frac{J}{T}\!\right)^{\!\!2} \nonumber
\\
&+&\, C^1_{3,0} \left(\!\frac{J}{T}\!\right)^{\!\!3}+...
\label{fe}
\end{eqnarray}
The coefficients $C_{b,0}^{1},\, b=1,2,3$, are given by \cite{ZT}
\begin{eqnarray}
C^1_{1,0} &=& 2A_+,\nonumber\\
C^1_{2,0} &=& 3A_+(1-2A_+)+3A_-,\\
C^1_{3,0} &=& {\textstyle{\frac{10}{3}}\displaystyle} A_+
(1-{\textstyle{\frac{27}{5}}\displaystyle} A_+ + 8 A_+^2)+8A_-(1-3A_+), \nonumber\label{Fcoef}
\end{eqnarray}
with
\begin{eqnarray}
C_0&=&  B_{0,D},\nonumber\\
A_+ &=& B_{D,0}/B_{0,D}^2,\nonumber\\
A_-&=&\exp(D/T)/B_{0,D}^3, \nonumber\\
B_{x,y}&=& 2\exp(x/T)\cosh(\mu_BgH/T)+\exp(y/T).\label{SP1c1}
\end{eqnarray}
\noindent

For later use, we also give the HTE free energy with in-plane  rhombic anisotropy $E$. 
If the external magnetic field is parallel to the $z$-axis, 
the chemical potentials in Eq.~(\ref{EQTM1}) become
\begin{equation}
\mu_1=h,\,\,\mu_2=D,\,\, \mu_3=-h,
\end{equation}
where $h=\sqrt{E^2+(\mu_Bg_{\parallel}H)^2}$.
In this case the function $B_{x,y}$ in Eq.~(\ref{SP1c1}) changes to
\begin{equation}
B_{x,y}= 2\exp(x/T)\cosh(h/T)+\exp(y/T).\label{SP1c2}
\end{equation}

On the other hand,
if we apply a perpendicular magnetic field to the Hamiltonian (\ref{Ham1}),  
the chemical potential terms in Eq.~(\ref{EQTM1}) are replaced by
\begin{eqnarray}
\mu_1&=&\frac{1}{2}(D-E+h'),\nonumber\\
\mu_2&=&E,\\ 
\mu_3&=&\frac{1}{2}(D-E-h'), \nonumber
\label{case3}
\end{eqnarray}
where $h'= \sqrt{(D+E)^2+4g_{\perp}^2\mu_B^2H_{a}^2}$ with $a=x$ or $y$ 
\cite{foot1}.
Subsequently, we have 
\begin{eqnarray}
C_0&=&   B_{(D-E)/2,D},\nonumber\\
A_+&=&B_{(D+E)/2,D-E}/B_{(D-E)/2,D}^2, \label{SP1c3}\\
A_-&=&\exp(D/T)/B_{(D-E)/2,D}^3, \nonumber
\end{eqnarray}
with now $B_{x,y}=2\exp(x/T)\cosh(h'/T)+\exp(y/T)$.

Eq.~(\ref{fe}) for the free energy $f(T,H)$ is our key result.
Physical properties such as the susceptibility, magnetization and the
specific heat follow in the usual way by differentiation.
We also use $f(T,H)$ to calculate the phase diagram for both
$T \simeq 0$ K and finite temperatures (see figure \ref{PD-NENC}).  
We find that considering up to 3rd order in $J/T$ is sufficient as higher
orders are negligibly small. 
This is in stark contrast to other series expansions which need many orders to 
accurately describe physical properties. 
This is mainly because here the coefficents are not just constants, but functions 
of the external model parameters, e.g., the magnetic field and the coupling strength.

\section{Spin-$1$ compounds}
 
\subsection{The compound NENC}

\begin{figure}
\begin{center}
\includegraphics[width=0.95\linewidth]{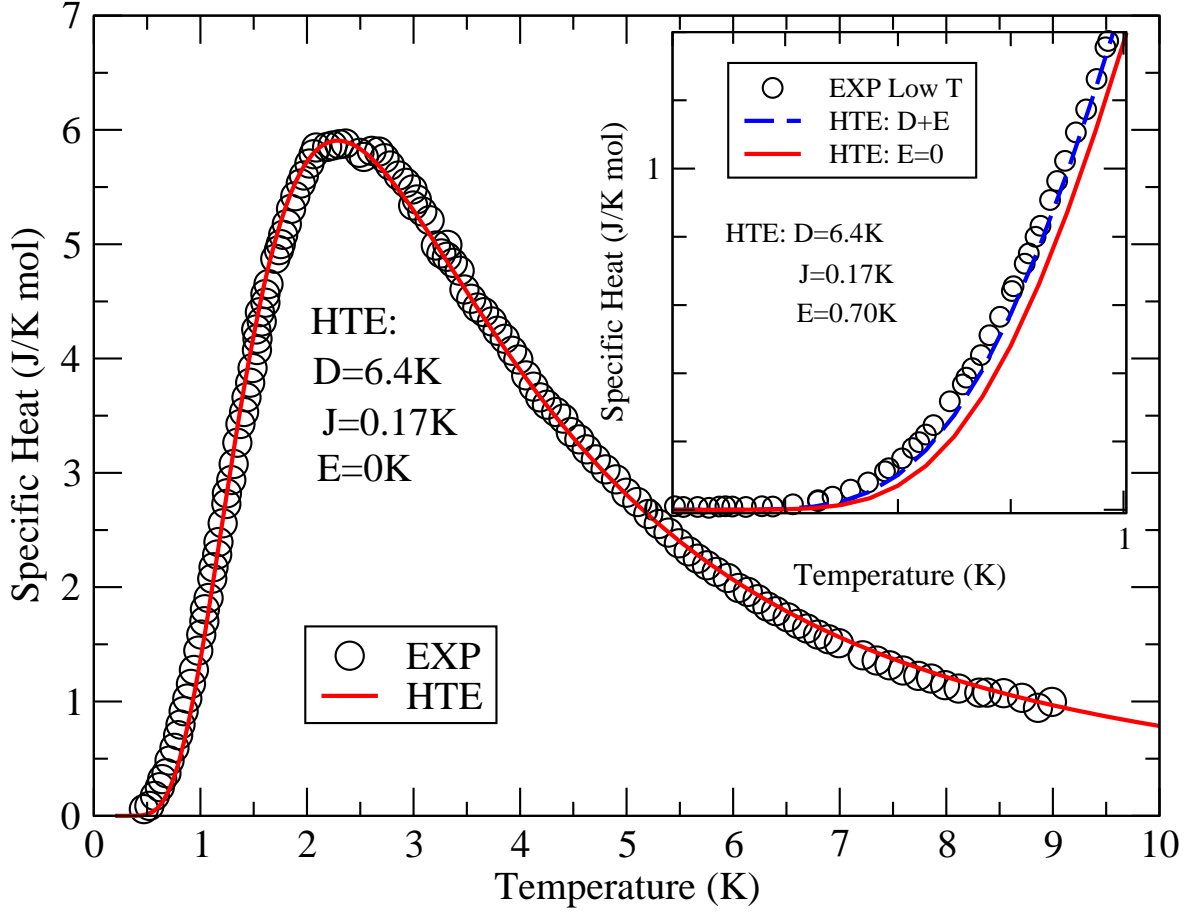}
\caption{
Comparison between theory and experiment \cite{NENC} for the
magnetic specific heat versus temperature of the compound NENC.
The conversion constant is $C_{{\rm HTE}} \approx 8 C_{{\rm EXP}}$ (J/mol-K).
The solid line denotes the specific heat  evaluated directly from the free energy (\ref{fe}) with the paramet
ers
$J=0.17$ K, $D=6.4$ K, $g=2.24$ and $\mu_B=0.672$ K/T.
The inset shows the low temperature specific heat.
Clearly the inclusion of in-plane rhombic anisotropy $E=0.7$ K (dashed line) gives a better
fit than without rhombic anisotropy (solid line).}
\label{FIGheat}
\end{center}
\end{figure}

It is known that antiferromagnetic spin-$1$ chains \cite{SP1C1,SP1C2}
with weak planar anisotropy can exhibit a non-magnetic gapped phase.
The large $D$ gapped phase has been observed in the compounds NENC,
NDPK and NBYC \cite{NENC,NBYC}. 
In these compounds the in-plane anisotropy $x^2-y^2$ breaks the $z^2$ symmetry 
and weakens the planar anisotropy. 
From experimental analysis, it was inferred that the
in-plane anisotropy $E$ in NENC \cite{NENC} is negligible in comparison
with the large $D$ single-ion anisotropy, where the Nickel(II) $z^2$ orbit 
along the c-axis forms a strong crystalline field. 
As a result the low temperature
physics is dominated by this strong crystalline field. 
The antiferromagnetic exchange interaction further lowers the energy but
its contribution to the ground state as well as the low-lying
excitations is minimal. 
As a consequence, the Hamiltonian (\ref{Ham1}) can be expected to describe this 
compound quite well.  
Experimentally, the specific heat was measured up to a temperature
around $10$ K in the absence of magnetic field \cite{NENC}.
A typical round peak
for short range ordering at $T\approx 2.4$ K is observed, see Figure~\ref{FIGheat}. 
An exponential decay is detected for temperatures
below approx $2.4$ K. 
Our calculated HTE specific heat for the Hamiltonian (\ref{Ham1}) 
with best visual fit constants $J=0.17$ K and $D=6.4$ K in the case 
(\ref{cp1}) (the solid line in Figure~\ref{FIGheat}) is in
excellent agreement with the experimental curve in the temperature
region $T>0.8$ K. 
In particular, the analytic result for the specific heat gives a
better fit with experimental data than the result from 
perturbation theory \cite{NENC}.  
For low temperatures (below $0.8$ K), paramagnetic impurities and 
a small rhombic distortion are the main reasons for the discrepancy.  
The inset of Figure~\ref{FIGheat} shows that the inclusion of a 
small rhombic anisotropy $E=0.7$ K gives a better fit for low temperatures 
than with $E=0$. 
However, at high temperature this rhombic anisotropy is negligible.

\begin{figure}
\begin{center}
\includegraphics[width=0.95\linewidth]{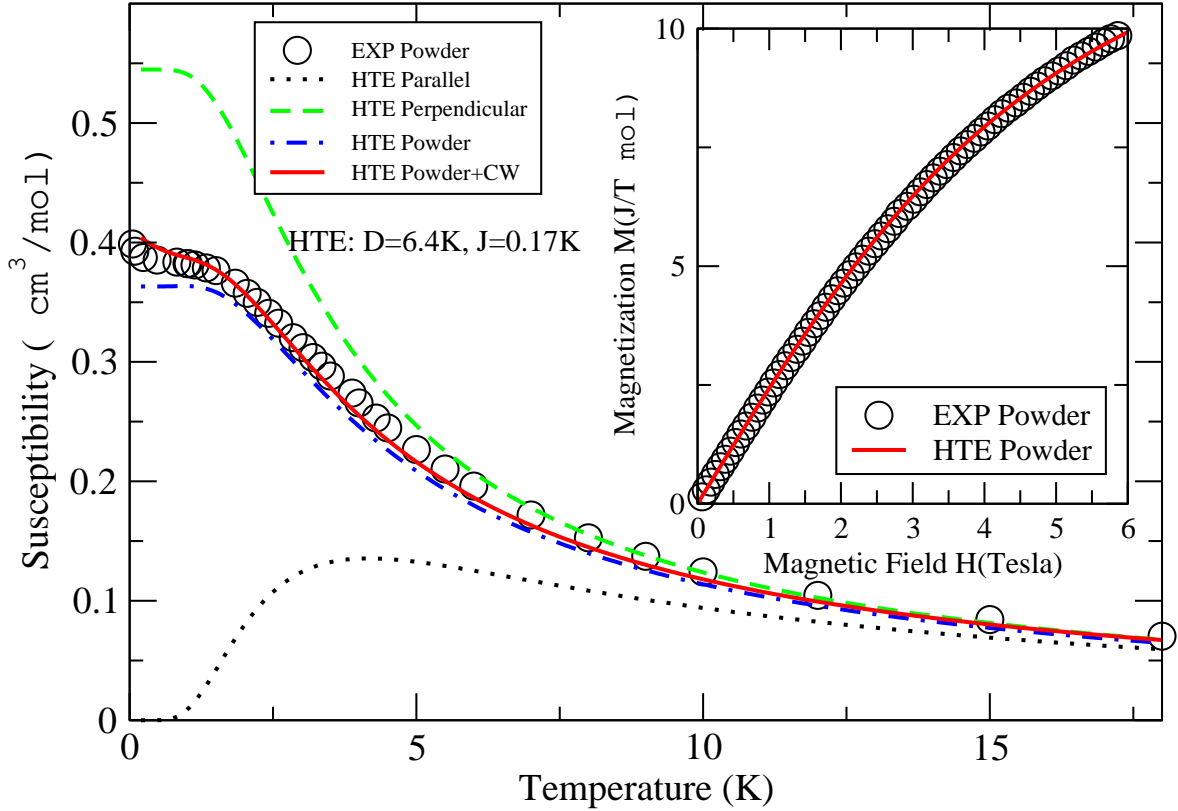}
\caption{
Comparison between theory and experiment \cite{sus} for the susceptibility
versus temperature of the compound NENC at $H=0.1$ mT.
The fitting curve (solid line) is obtained via the empirical relation
$\chi _{{\rm Powder}} \approx \frac{1}{3}\chi_{\parallel}+\frac{2}{3}\chi_{\perp}$
together with a Curie-Weiss (CW) contribution.
The inset shows the comparison between theory and experiment \cite{NENC} for the
magnetization versus magnetic field of NENC at the temperature $T=4.27K$.
A good fit for both the susceptibility and magnetization suggests the coupling
constants $J=0.17$ K, $D=6.4$ K, $g_{\perp}= 2.18$ and $g_{\parallel}=2.24$.
The conversion constants are
$\chi_{{\rm HTE}}\approx 0.8123 \chi_{{\rm EXP}}$(cgs/mol) and
$M_{{\rm HTE}}\approx 8.5 M_{{\rm EXP}}$ ($10^3$ cgs/mol).
}
\label{FIGSZ}
\end{center}
\end{figure}

As far as we know, the susceptibility was measured only for powdered samples of 
this compound. 
Moreover, the experimental susceptibility of NENC was studied only in the 
temperature range 50 mK - 18 K under a static magnetic field $H=0.1$ mT.  
From the data shown in Ref.~\cite{NENC} we cannot accurately
estimate the contributions for the Curie-Weiss term and the para\-magnetic 
impurity.
In Figure~\ref{FIGSZ} we present our theoretical curves
for the susceptibility with parallel and perpendicular field evaluated from
the free energy associated with different chemical potentials.  
A susceptibility estimation for powdered samples using 
$\chi _{{\rm Powder}} \approx \frac{1}{3}\chi_{\parallel}+\frac{2}{3}\chi_{\perp}$ 
\cite{Carlin} does not fit the experimental data 
very well at low temperatures due to the Curie-Weiss contribution and 
paramagnetic impurities. 
A visual fit with the experimental susceptibility suggests that the contribution from the 
Curie-Weiss term is not negligible. 
We find that our theoretical susceptibility $\chi _{{\rm Powder}}$ for powder 
with a Curie-Weiss contribution $c/(T-\theta)$ gives a satisfactory
agreement with the experimental curves, where $c \approx 0.045 $
cm$^{3}$ K/mol and $\theta \approx -0.9$ K.  
This fit suggests the values $J=0.17$ K and $D=6.4$ K, with $g_{\perp}= 2.18$ and $g_{\parallel}=2.24 $.  
From the TBA analysis we find an energy gap $\Delta \approx 5.72 $ K with a parallel
external magnetic field at zero temperature for these coupling constants.
The typical antiferromagnetic behaviour of the susceptibility with a parallel
magnetic field to the axis of quantization follows from our results.
This is in accordance with the behaviour of the specific
heat given in Figure~\ref{FIGheat}.
The inset of Figure~\ref{FIGSZ} shows the magnetization of a
powdered sample at $T=4.27$ K. It is obvious that the singlet is
supressed by the temperature. 
Fitting suggests the empirical relation 
$M_{\rm Powder} \approx \frac{1}{3}M_{\parallel}+\frac{2}{3}M_{\perp}$
for the powdered magnetization with the same constants as before.  
Here $M_{\parallel}$ and $M_{\perp}$ denote the
magnetization with the field parallel and perpendicular to the axis
of quantization.

\subsection{The compound NBYC}

\begin{figure}
\begin{center}
\includegraphics[width=0.95\linewidth]{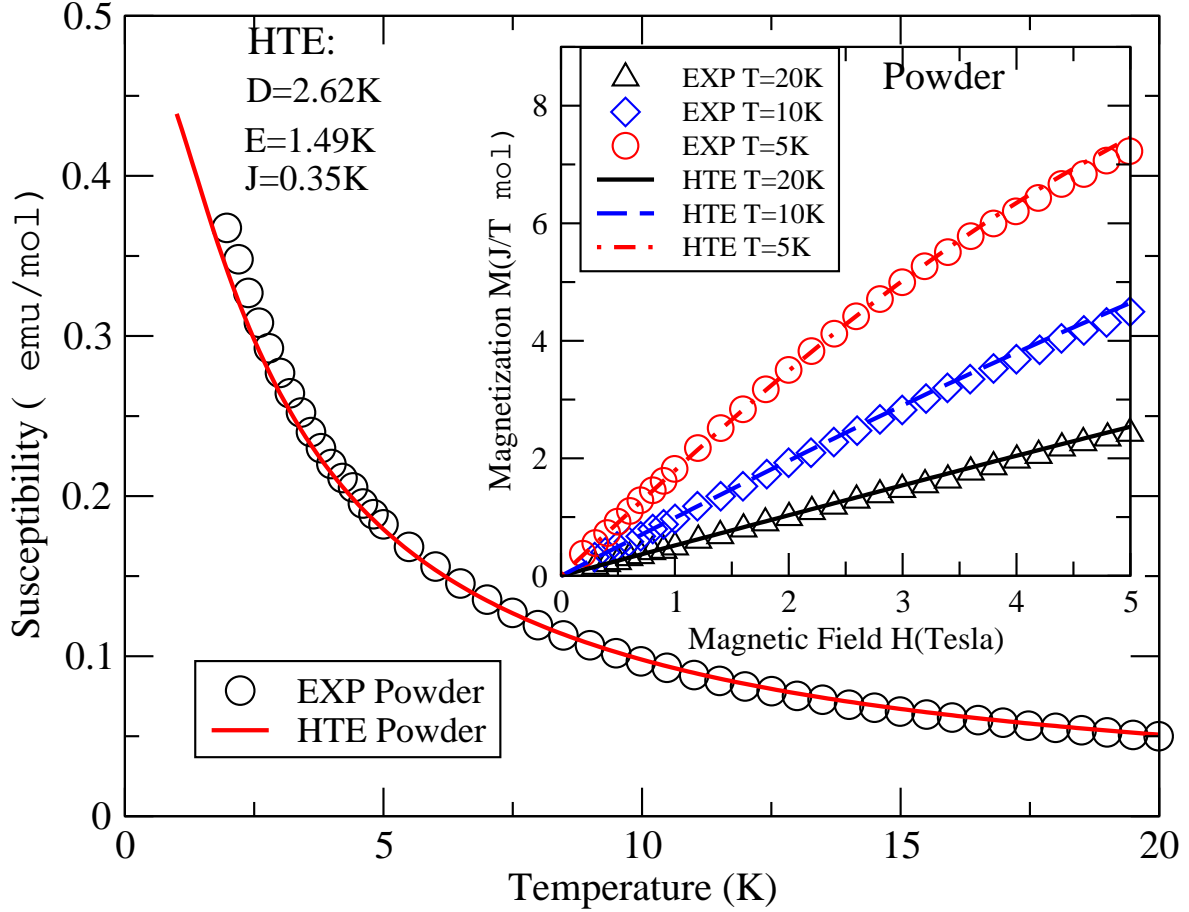}
\caption{
Comparison between theory and experiment \cite{NBYC} for the
susceptibility versus temperature of the compound  NBYC.
The conversion constants are the same as for NENC.
The solid line is the susceptibility for the powdered samples with coupling
constants $D=2.62$ K, $E=1.49$ K and
$J=0.35 $ K, with $g_{\parallel}=g_{\perp}=2.05$.
The small discrepancy at low temperature might be attributed to a Curie-Weiss contribution.
The inset shows the magnetizations for powdered samples at $5$ K, $10$ K and $20$ K.
In each case the theoretical results verify
the existence of weak exchange coupling and in-plane rhombic anisotropy, with
a strong single-ion anisotropy.
}
\label{FIGNBYC}
\end{center}
\end{figure}

We now turn to the properties of the compound NBYC, which has also been
experimentally investigated \cite{NBYC}.  
In particular, in-plane anisotropy $E$ and a large anisotropy $D$ are present,
suggesting that the model Hamiltonian (\ref{Ham1}) may again be a 
good microscopic model for this type of compound. 
Theoretical studies  based on strong-coupling
expansion methods \cite{Spathis} suggest that the anisotropy of this
compound might lie in the vicinity of the boundary between the Haldane
and field-induced gapped phases \cite{NBYC}. 
However, due to the validity of the strong-coupling expansion method, the fits for
specific heat, susceptibility and magnetization become increasingly inconsistent
with each other as the rhombic anisotropy increases.  
Figure ~\ref{FIGNBYC} presents the susceptibility for this compound.
The theoretical susceptibility curve for the powdered sample is
evaluated from the free energy (\ref{fe}) with parallel and perpendicular fields
(see Eq.~(\ref{case3})) via the empirical formula 
$\chi_{{\rm Powder}} \approx \frac{1}{3}\chi_{\parallel}+\frac{2}{3}\chi_{\perp}$.
A good fit for the susceptibility suggests the values $D=2.62$ K, $E=1.49$ K,
$J=0.35 $ K, with $g_{\parallel}=g_{\perp}=2.05$.  A small discrepancy
at low temperature can be attributed to a Curie-Weiss contribution term. 
From the TBA analysis we conclude that the ground state is gapless.

The inset of Figure~\ref{FIGNBYC} shows the magnetization for
powdered samples at $5$ K, $10$ K and $20$ K.  Again our theoretical
curves are evaluated using the empirical relation
$M_{\rm Powder} \approx \frac{1}{3}M_{\parallel}+\frac{2}{3}M_{\perp}$
for the powdered magnetization.
An overall agreement
in magnetization for different temperatures gives a consistent parameter
setting for the susceptibility. The singlet state  is now
supressed by the in-plane rhombic anisotropy and the temperature.

\begin{figure}
\begin{center}
\includegraphics[width=0.95\linewidth]{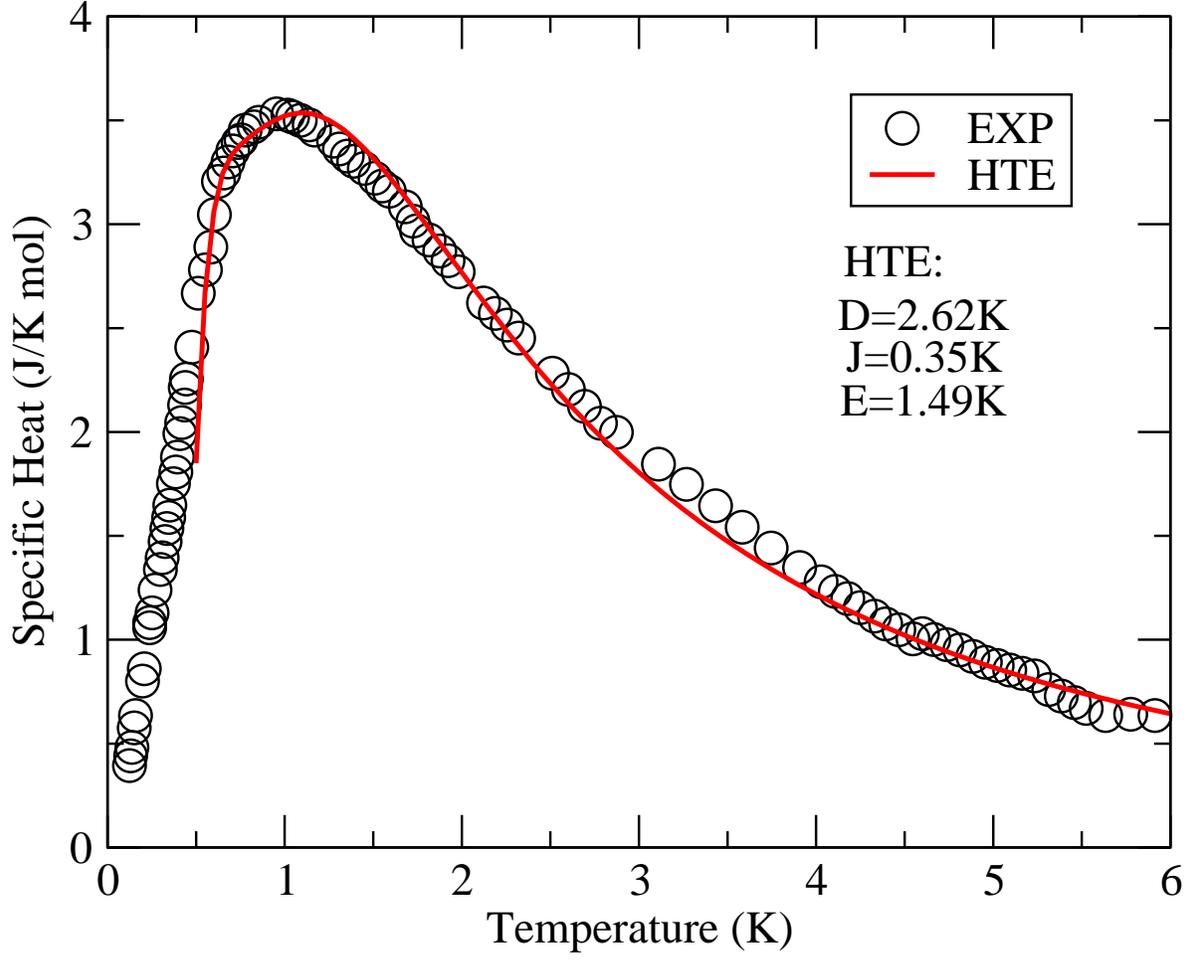} 
\caption{
Comparison between theory and experiment \cite{NBYC} for the
magnetic specific heat versus temperature of the compound  NBYC.
The conversion constant is $C_{{\rm HTE}} \approx 10 C_{{\rm EXP}}$ (J/mol-K).
The solid line denotes the specific heat at $H=0.1$ mT evaluated directly
from the free energy (\ref{fe})
with the  same parameters as in Figure~\ref{FIGNBYC}.
}
\label{FIGNBYCheat}
\end{center}
\end{figure}

The specific heat was measured up to a
temperature of $6$ K in absence of magnetic  field
\cite{NBYC}. The theoretical  specific heat evaluated from the model Hamiltonian
(\ref{Ham1}) (the solid line in Figure \ref{FIGNBYCheat}),
with the same parameters used before, is in good agreement with the experimental curve
in the temperature region $0.5$ K to $6$ K. For temperatures below $0.5$ K
the high temperature expansion does not converge and thus
cannot provide valid predictions.

\section{Conclusion}

We have investigated the thermal and magnetic properties of spin-$1$ compounds with large 
single-ion anisotropy, such as NENC and NYBC, via the thermodynamic Bethe Ansatz 
and the high temperature expansion for the integrable model (1).
Excellent agreement was found with the experimental magnetic properties of these compounds \cite{foot2}.
The large single-ion anisotropy results in a nondegenerate singlet ground
state which is different from the valence bond solid Haldane phase. The
in-plane anisotropy weakens the energy gap.

\begin{figure}[t]
\vspace{10mm}
\includegraphics[width=0.95\linewidth]{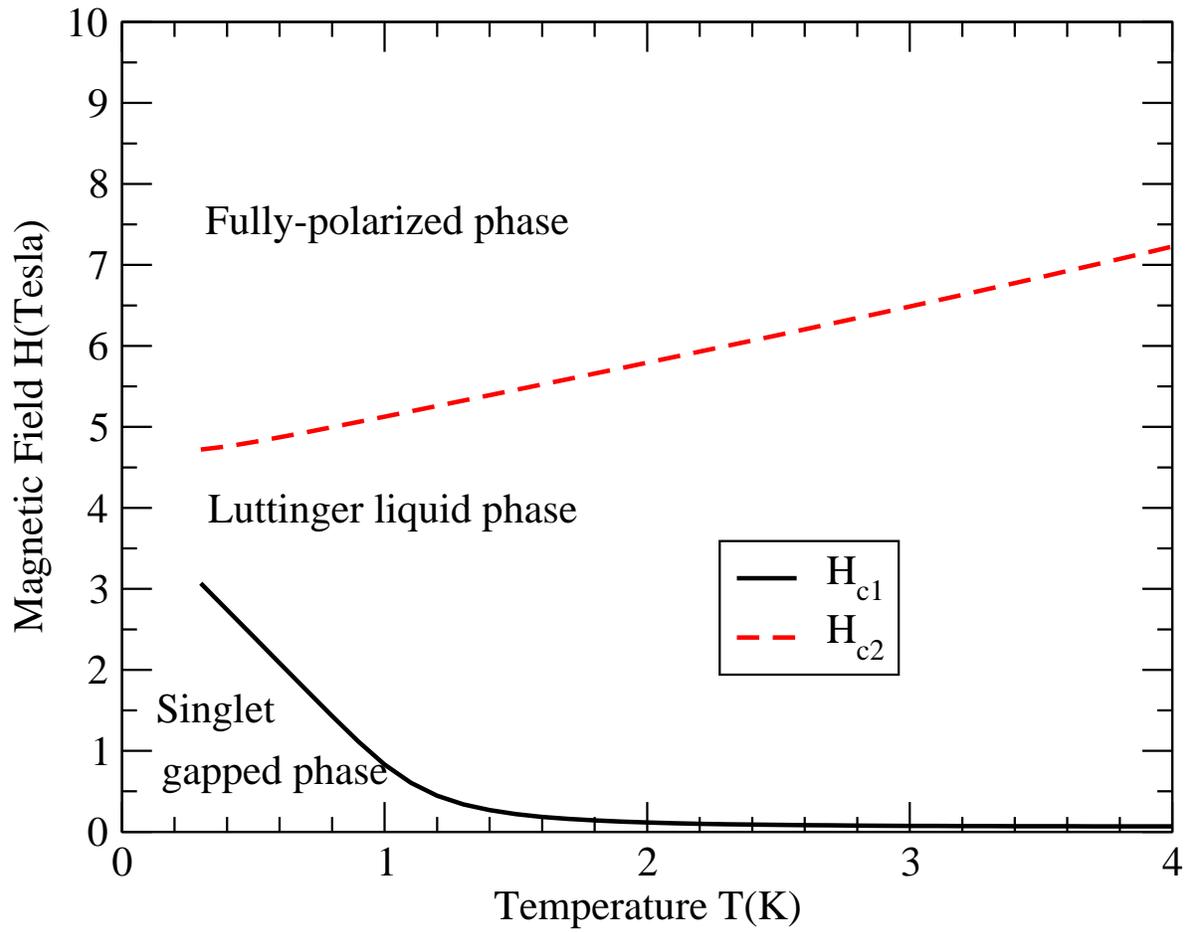}
\caption{
Phase diagram for the compound NENC with parameters
$J=0.17$ K, $D=6.4$ K, $g=2.24$ and $\mu_B=0.672$ K/T.
}
\label{PD-NENC}
\end{figure}

Finally, we give the full phase diagram of the compound NENC in Fig.~\ref{PD-NENC}. 
We see that the gapped phase is quickly exhausted as the temperature increases. 
The magnetic ordered Luttinger liquid phase
lies between the curves defined by $H_{c1}$ and $H_{c2}$. 
The ferromagnetic polarized phase is above the $H_{c2}$ curve.  
The intersection of the critical curves and the $H$-axis indicates
the estimated values $H_{c1} \approx 3.8$ T and $H_{c2}\approx 4.7$ T, 
which coincide with the TBA results at $T=0$ K discussed in section
\ref{sec:TBA}.
However, for the case where the in-plane anisotropy $E\neq 0$, the
critical behaviour is different from the phase diagram of Fig.~\ref{PD-NENC}.
In this case the fully-polarized phase appears for $H>>H_{c2}$
because the in-plane anisotropy mixes the doublet components 
$\mid \!\! S^z=\pm 1 \rangle$.
We anticipate that the exact results for the susceptibility and the magnetization of 
the powdered samples as well as for the compounds with parallel and perpendicular magnetic fields
may find widespread use in the study of their magnetic properties and for identifying
the quantum effects resulting from single-ion anisotropy. 
Our analytic approach via the Hamiltonian (1) may thus 
describe the thermal and magnetic properties of other compounds, such as 
NDPK \cite{NENC,sus}, NiSnCl$_6\cdot 6$H$_2$O \cite{PRB3488},
 [Ni(C$_5$H$_5$NO)$_6$](ClO$_4$)$_2$ \cite{PRB3523},
and Ni(NO$_3$)$_2\cdot 6$H$_2$O \cite{PRB4009}.

\noindent
{\em Acknowledgements.} This work has been supported by the Australian Research Council. 
N. Oelkers also thanks DAAD for financial support.
We thank Z. Tsuboi, A. Foerster and H.-Q. Zhou for helpful discussions.
We also thank M. Orend\'{a}\u{c} for providing us with experimental results and helpful discussions.


\begin{thebibliography}{99}
\bibitem{foot1} Equivallently, we can consider the external field
parallel to the $z$ axis, whereas the crystalline field and
orbital splitting are along $x$ or $y$ axis.
\bibitem{foot2} We find similar agreement for the compound NDPK.
\end{thebibliography}
\end{document}